\documentclass[%
 aip,
 jmp,%
amsmath,amssymb,
preprint,%
]{revtex4-1}

\usepackage{graphicx}
\usepackage{dcolumn}
\usepackage{bm}
\usepackage{epstopdf}
\usepackage{color}
\usepackage{ulem}
\usepackage{amsthm,amsmath,amssymb}
\usepackage{mathrsfs}
\usepackage{indentfirst}
\usepackage{caption}
\usepackage{hyperref}

\hypersetup{
  colorlinks   = true, 
  urlcolor     = blue, 
  linkcolor    = blue, 
  citecolor   = blue 
}

\begin{document}
\title{Proton Tunneling in a Two-Dimensional Potential Energy Surface with
a Non-linear System-Bath Interaction: Thermal Suppression of Reaction Rate}


\author{Jiaji Zhang}
\email{zhang.jiaji.84e@st.kyoto-u.ac.jp}
\affiliation{Department of Chemistry, Graduate School of Science, Kyoto University, Kyoto 606-8502, Japan}

\author{Raffaele Borrelli}
\email{raffaele.borrelli@unito.it}
\affiliation{DISAFA, University of Torino,  Largo Paolo Braccini 2, I-10095 Grugliasco, Italy}

\author{Yoshitaka Tanimura}
\email{tanimura.yoshitaka.5w@kyoto-u.jp}
\affiliation{Department of Chemistry, Graduate School of Science, Kyoto University, Kyoto 606-8502, Japan}
\date{\today}

\begin{abstract}
We consider a proton-transfer (PT) system described by a proton-transfer reaction (PTR) coordinate and a rate promoting vibrational (RPV) coordinate interacting with a non-Markovian heat-bath. While dynamics of PT processes has been widely discussed using two-dimensional (2D) potential energy surfaces (PES), the role of the heat-bath, in particular, for a realistic form of the system-bath interaction has not been well explored. Previous studies are largely based on one-dimensional model and linear-linear (LL) system-bath interaction. In the present study, we introduce an exponential-linear (EL) system-bath interaction, which is derived from the analysis of a PTR-PRV system in a realistic situation. This interaction mainly causes vibrational dephasing in the PTR mode and population relaxation in the RPV mode. Numerical simulations were carried out using hierarchy equations of motion approach. We analyze the role of the heat-bath interaction on the chemical reaction rate as a function of the system-bath coupling strength at different temperature and for different values of the bath correlation time. A prominent feature of the present result is that while the reaction rate predicted from classical and quantum Kramers theory increases as the temperature increases, the present EL interaction model exhibits opposite temperature dependence. Kramers turn-over profile of the reaction rate as a function of the system-bath coupling is also suppressed in the present EL model turning into a plateau-like curve for larger system-bath interaction strength. Such features arise from the interplay of the vibrational dephasing process in the PTR mode and the population relaxation process in the RPV mode.
\end{abstract}
                         
\keywords{Proton transfer, Two-dimensional vibrational spectroscopies, Hierarchical Equations of motion}

\maketitle
\section{Introduction}

Proton transfer (PT) processes play an essential role in many types of biological and functional molecular materials and determines the efficiency of the mechanisms for utilizing chemical and solar energies.\cite{Wolyness1981,Miller1983,Miller1989, HanggiRMP90,SobolewskiPNAS05,LanPNAS08,LiuPCCP13} One of the central questions in the PT problem is how thermal effects, most typically fluctuation and dissipation that arise from a molecular environment (a heat-bath), influence the efficiency of the PT reaction. 

This problem has been a subject of open quantum dynamics theory, because the PT involves a deep tunneling process that requires full quantum mechanical description of the system, and because the PT system undergoes an irreversible dynamics through interactions with an environment, for example a solvent, a protein scaffold or a nanostructured material, in a non-Markovian and non-perturbative manner.

Several theoretical approaches have been developed to account for quantum effects of chemical reactions,
\cite{CEPenJPCA2019,KawaTachiJCTC2014,VothJPC1993,BilGunCPC1997,AntAboSchwJCP2004} 
but the capability to treat deep tunneling problems from such approaches is limited. 
Moreover, many investigations have been limited to systems defined in a one-dimensional configuration space linearly coupled with the Markovian heat-bath.
\cite{PRSA2018,PAVOLCP1986,AntSchwPNAS1997,HofCotThoJCP2017,SoudHSchiJCP2015,
DoStaMavCP2001,BasBorVuiPCCP2013,WSMiller1998} 
In realistic situations, the potential energy surface (PES) of the PT system is defined at least in a two-dimensional configuration space, which consists of a proton-transfer reaction (PTR) mode and a rate promoting vibrational (RPV) mode that comprises mostly a stretching vibration of the hydrogen-bridged moieties. 
Furthermore, the system-bath interaction can be highly non-linear, which causes strong vibrational dephasing effect on the system dynamics.

While the importance of the RPV modes have been discussed intensively, \cite{DzierSchJPCL2016,SchrSchwBioChem2018,NatChem2016} majority of treatments have ignored quantum mechanically entangled environmental effects, assuming static potential or employing adiabatic approximation under external perturbation. 
\cite{HynCPL1989,BorHynCP1993,PetJCTC2010,AntSchwJCP1998,JSSutJRSI2008}
In realistic situation, however, proton transfer processes are controlled by transition distance and energy barrier height as a function of RPV coordinate which is further coupled to the environmental degrees of freedom. The key aspect is that the system-bath interaction in the PT process is non-Markovian, and has to be treated non-perturbatively at full quantum mechanical level, while most of the existing theories cannot treat this effect accurately. 
\cite{Shi2009PT,Shi2011PT,HanggiRMP90,MPKDNatCom2018,MaMoSaJPC2013,PNAS2011}

Meanwhile, from the investigation of exciton transfer processes in photosynthetic antenna systems, it has been realized that the non-Markovian and non-perturbative system-bath interaction in quantum regime play a central role in determining the efficiency of the excition transition rate. \cite{IshiFlemJCP09, SakaTaniJPCL2017,JZSKJPCB2011,BorrelliGelinSR2017} 
Similar to the exciton transfer problems, it is expected that the quantum nature of the system-bath interaction is also essential in determining the efficiency of the PT reaction. 

In this paper, we give an extension of PT theory focusing on the role of a realistic system-bath interaction to help further development of the investigation in this area. We consider a PT system, which is described as a two-dimensional potential coupled to the environment at finite temperature. More specifically, we consider an exponential form of the system-bath interaction that includes a RPV-bath coupling term to describe thermal effect of the RPV mode. Employing the exact hierarchal equations of motion (HEOM) theory, we numerically simulate the PT process for different system-bath coupling strengths, noise correlation time, and at different temperature.

The organization of the paper is as follows. In Sec.\ref{sec:Theory}, we present a model system for proton transfer process. Then we show how the chemical reaction rate can be calculated from the HEOM on the basis of the linear response theory. In Sec.\ref{sec:result}, we present the numerical results and discussion, and the conclusion is given in Sec.\ref{sec:Conclusion}.

\section{Theory}
\label{sec:Theory}
\subsection{The system-bath Hamiltonian}
\label{sec:SB}

\begin{figure}[t]
\centering
\includegraphics[width=0.9\textwidth]{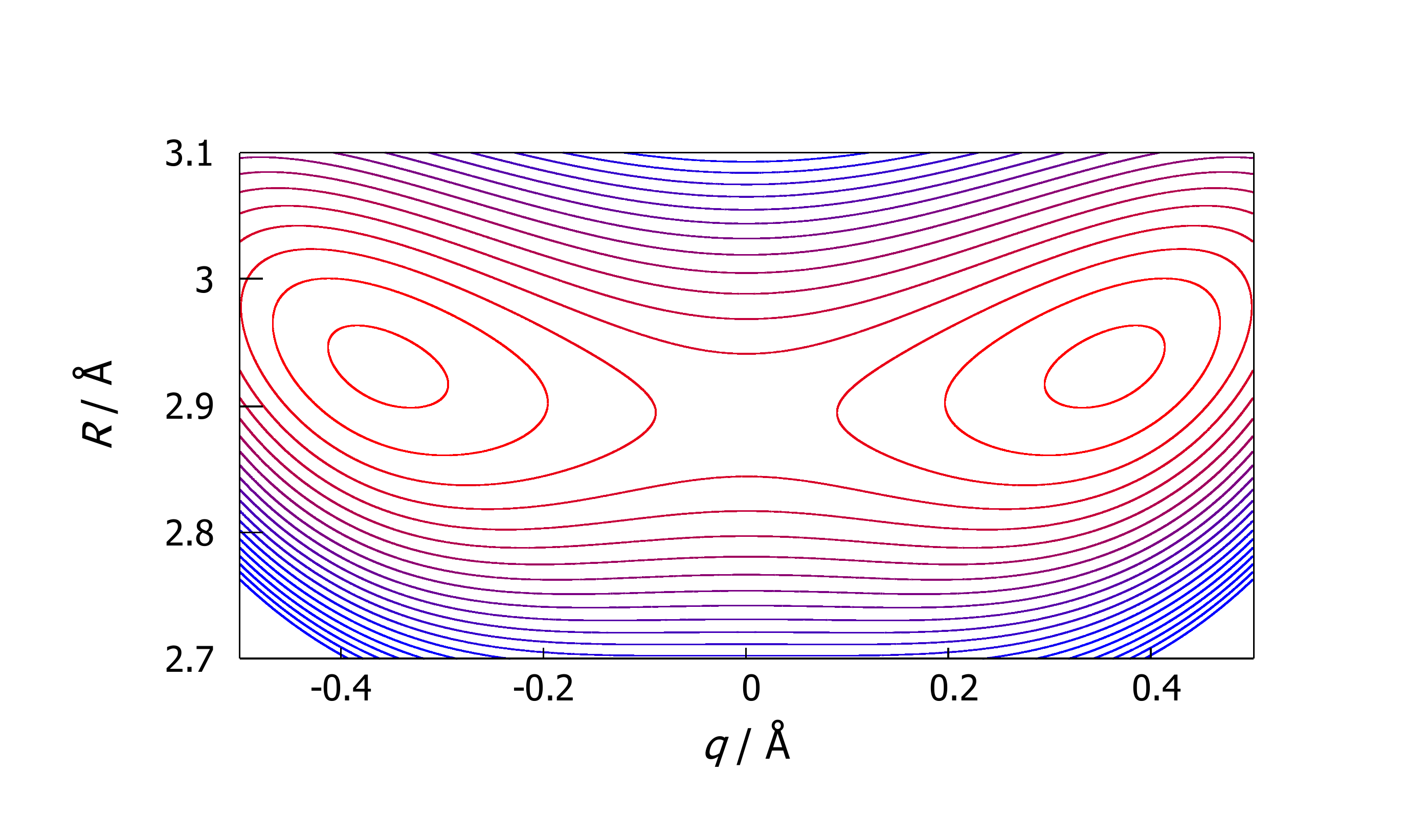}
\caption{The two-dimensional potential energy surface $U_0(\hat q, \hat R)$ is depicted. The parameters of the  model are reported in table \ref{table.parameter}.  Contour lines are drawn every 500 ${\rm cm}^{-1}$.}
\label{fig.contourmap}
\end{figure}

We consider a proton transfer (PT) system described by a PTR coordinate and rate promoting vibrational (RPV) modes. While we can treat more than three-dimensional configuration space using the energy eigenstate representations of the system, for the sake of conciseness, here we consider the two-dimensional case as a minimal model for PT process. In our model the potential energy surface (PES) is modulated by an environment (bath) representing the other non-RPV vibrational modes which might represent any type of solvation or protein modes. 

The system Hamiltonian is then expressed in the form
\begin{align}
  \hat{H}_{S}(\{\hat x_j\}) &= \frac{\hat p^2}{2m}+\frac{\hat P^2}{2M} + U(\hat q, \hat R; \{\hat x_j\}).
  \label{eq:system-Hamiltonian}
\end{align}
Here $\hat q$, $\hat p$, and $m$ represent the coordinate, momentum, and mass of the PTR coordinate, $\hat R$, $\hat P$, $M$ those of the RPV mode, and $\{\hat x_j\}$ are the coordinates of the bath modes. 
In the above form $H_S$ includes the modulation of the system parameters induced by the bath coordinates.

The potential $U(\hat q, \hat R; \{\hat x_j\})$ is written as
\begin{align}
  U(\hat q, \hat R; \{\hat x_j\}) &= D_0 \left( 1 - e^{ -\alpha(\{\hat x_j\}) ( {\hat R}/{2} -{\hat q} -q_e)} \right)^{2} \nonumber \\
  &+ D_0 \left( 1 - e^{ -\alpha(\{\hat x_j\}) ( {\hat R}/{2} + {\hat q} -q_e)} \right)^{2}  + \frac{1}{2} D_k(\hat R - R_e)^{2},
  \label{eq:potential}
\end{align}
where $D_0$ is the dissociation energy, $q_e$ is the equilibrium distance of the proton, $\alpha(\{x_j\})$ is the curvature of Morse potential which depends on the 
the bath degrees of freedom $\{x_j\}$ (see below),
  $D_k$ is the force constant for the RPV mode, and $R_e$ is the equilibrium distance between the donor and acceptor without the presence of the proton.

This form of the two-dimensional potential of an hydrogen bridged system has been successfully used in the past to model hydrogen bond dynamics and spectroscopy.\cite{SatoIwata,AlfanoJPCA02,LippincottJCP55} 
The unique feature of the present model is that the effect of the environment is included via the dependence of the anharmonicity parameter $\alpha$ on the bath coordinates. The rationale behind this choice is that in the above potential energy both the height of the barrier for the proton tunneling, and the vibrational frequency of the H-bond depend linearly on the $\alpha$ parameter.
Furthermore, as we will shortly demonstrate, this type of coupling enable us to include a linear coupling between the RPV mode and the bath.

The $\alpha$ aparmeter of the potential is considered to be linearly depend on the coordinates $\{x_j\}$ as $\alpha(\{x_j\}) =  \alpha_0 + \alpha^{(1)}\sum_{j} g_j \hat x_j$, where $\alpha_0$ and $\alpha^{(1)}$ are the constants that represent the curvature of the system-potential and the strength of the coupling between the system and the environment. Finally, by expanding the PES in power series around $\{x_j\}=0$ we obtain 
\begin{align}
  U(\hat q, \hat R; \{x_j\}) &= U_0(\hat q, \hat R) + V(\hat q, \hat R ) \sum_{j}g_j
\hat x_j ,
 \label{eq:potential2}
\end{align}
where $ U_0(\hat q, \hat R) \equiv U(\hat q, \hat R; \{\hat x_j=0\})$ is the hydrogen bridge potential without the bath interaction (see Fig. \ref{fig.contourmap}) and the interaction potential $V(\hat q, \hat R )$ is expressed as 
\begin{align}
 V(\hat q, \hat R )  &=V_0 ({\hat R}  - 2q_e)- V_0 e^{ -\alpha_0 ( {\hat R}/{2} -q_e)}\left[ ({\hat R} - 2q_e ) \cosh(\alpha_0 \hat q) - 2 \hat q \sinh(\alpha_0 \hat q) \right],
\label{eq:V}
\end{align}
where $V_0 = \alpha^{(1)}D_0$. Following the standard approach of the theory of open quantum systems, we assume that the bath modes are represented by a set of harmonic oscillators, with the $j$-th bath oscillator possessing the frequency $\omega_{j}$, mass $m_{j}$, position coordinate $\hat{x}_{j}$ and momentum $\hat{p}_{j}$. The total Hamiltonian is then expressed as
\begin{align}
\hat H_{\rm tot} &= \hat H_{\rm S} + \sum_{j} {\left[ \frac{\hat p_j^2 }{2m_j } + \frac{1}{2}m_j \omega _j^2 \left( \hat x_j  -\frac{ g_j  \hat V(\hat q, \hat R ) }{m_j \omega _j^2} \right)^{2} \right] },
\label{eq:SpinBoson}
\end{align}
where $\hat H_{S}$ is the system Hamiltonian for the potential $U_0(\hat q, \hat R)$ and we included the counter term in the system-bath interaction.\cite{TanimuraPRA91,TanimuraJCP92,KatoJPCB13} Hereafter, we refer to $\hat V(\hat q, \hat R ) \sum_{j}g_j \hat x_j $ as the exponential-linear (EL) interaction due to the form of the operator $\hat V$. The EL interaction mainly modulates the barrier height of the proton potential, which can be used to investigate  non-polarizable solvent effect of a realistic PTR process.\cite{HynJPCB2016,SmithSci1992}
The heat bath is characterized by the spectral distribution function (SDF), defined by 
$J(\omega) \equiv \sum {\hbar g_{j}^2} \delta(\omega-\omega_{j})/{2m_{j} \omega_{j} }$, and the inverse temperature, $\beta \equiv 1/k_{\mathrm{B}}T$, where $k_\mathrm{B}$ is the Boltzmann constant. 
By adjusting the form of the SDF, the properties of the local environment, for example the solvates and protein molecules, can be modeled. 
Typically, the SDF is estimated from linear and nonlinear infrared and Raman spectra, both experimentally\cite{IR_exp1,Palese96, Pullerits20} and numerically.\cite{Renger06, KramerAspu13, KramerAspu14, Coker2016,Ueno2020}

The main features of the EL interaction can be illustrated by expanding $V(\hat q, \hat R )$ in the Taylor series with respect to $\hat q$ and $\hat R$ as
\begin{align}
 \hat{V}(\hat{q}, \hat{R}) =\hat{V}^{(2,0)} \hat{q}^2  + \hat{V}^{(0,1)}\hat{R} + \hat{V}^{(2,1)}\hat{q}^2 \hat{R}  +  \dots ,
\label{eq:Vtaylor}
\end{align}
where{$\hat{V}^{(k,l)}= {\partial^{k+l} \hat{V}}/{\partial \hat{q}^k \partial \hat{R}^l}|_{\hat{q}=\hat{R}=0}$. It should be noted that, due to the symmetry of the potential along the proton coordinate,  $V(\hat q, \hat R )$ is an even function of the proton coordinate $\hat{q}$. Thus the leading order of the EL model for the PTR mode is the square-linear (SL) interaction, expressed as $\hat q^2 \sum_j g_j x_j$, while that for the RPV mode is the linear-linear (LL) interaction, expressed as $\hat R \sum_j g_j x_j$. The distinct feature of the present model arises from this SL interaction in the PTR mode, while most of the system-bath models employed in the former investigations were assumed to be the LL system-bath interaction. 

Although the SL interaction has been used to analyze the effects of vibrational dephasing in linear spectroscopy \cite{Lynden-BellMP77,OxtobyJCP78,VelskoJCP80,MarksJCP80} as well as
in ultrafast nonlinear spectrocopies,\cite{OkumuraPRE97, SakuraiJPCA11} the relevance of such interaction has not been properly addressed in the PTR problem. We notice that our model also includes a linear coupling of the RPV mode which has been found to be quite significant in molecular dynamics simulation of hydrogen atom transfer.\cite{AthokpamJCP18}

In order to calculate the reaction rate, we employ the eigenstates representation for both $\hat H_S$ and $\hat V(\hat q, \hat R)$. (See Appendix \ref{sec:Uqr}.)

\subsection{The reduced hierarchical equations of motion formalism}
\label{sec:HEOM}

In the framework of the system-bath Hamiltonian, the characteristic feature of the environment is determined by the choice of the SDF. A typically employed SDF for a molecular environment is the Drude SDF, \cite{Tanimura89A,TanimuraPRA90,TanimuraPRA91,TanimuraJCP92, KatoJPCB13,IshizakiJPSJ05, tanimura2006,Tanimura2014,Tanimura2015, BorrelliJCP19} and the Brownian SDF\cite{TanimuraMukamelJPSJ94,TanakaJPSJ09}, the Ohmic SDF \cite{Ikeda2019Ohmic}, and their combinations.\cite{TanimuraJCP12, KramerAspu13, Shi2014SpecralDecom} 

Here, we consider the Drude SDF defined by 
\begin{equation}
 J(\omega) = \frac{\hbar \zeta}{\pi}\frac{\gamma^2\omega}{\gamma^2+\omega^2},
\label{eq:JDrude}
\end{equation}
where the parameter $\gamma$ represents the width of the spectral distribution of the collective bath modes and is the reciprocal of the noise correlation time induced by the heat bath, $\tau_c$ = 1/$\gamma$. The parameter $\zeta$ is the system-bath coupling strength, which represents the magnitude of fluctuations and  dissipations.
The heat bath effect is characterized by the noise correlation function, $C(t) \equiv \langle {\hat X}_k(t) {\hat X}(0)\rangle$, which can be further expressed as a linear combination of exponential functions and a delta function,
 $C(t) =\sum_{k} (c'_{k}+i c''_{k}) \gamma_{k}e^{-\gamma_{k}\vert t \vert}+2c_{\delta}\cdot \delta_t$. 
 \cite{Tanimura89A,TanimuraPRA90,IshizakiJPSJ05,tanimura2006}
 Then, using path integral method, reduced hierarchy equations of motion (HEOM) can be derived as\cite{Tanimura89A,TanimuraPRA90,IshizakiJPSJ05, tanimura2006,Tanimura2014,Tanimura2015,BorrelliJCP19}
\begin{align}
\frac{\partial}{\partial t} \hat{\rho}_{\vec{n}}(t)
= &\, - \left[ \frac{i}{\hbar} \mathcal{\hat L}_S 
      + \sum_{k=1}^K n_{k} \gamma_{k}+c_\delta \hat{\Phi}^2 \right]
      \hat{\rho}_{\vec{n}}(t)
\notag \\
  &\, - \hat{\Phi}\sum_{k=1}^K \hat{\rho}_{\vec{n} + \vec{e}_k}(t)
  - \sum_{k=1}^K n_{k} \hat{\Theta}_{k}
      \hat{\rho}_{\vec{n} - \vec{e}_k}(t),
\label{eq:HEOM}
\end{align}
where we introduce the operators $\mathcal{\hat L}_S \hat{\rho}\equiv[\hat{H}_S, \hat{\rho}]$, $\hat{\Phi}\hat{\rho}\equiv(i/\hbar)[\hat{V}, \hat{\rho}]$, $\hat{\Psi}\hat{\rho}\equiv(1/\hbar)\{ \hat{V}, \hat{\rho} \}$, and $\hat{\Theta}_{k} \equiv c'_{k} \hat{\Phi} - c''_{k} \hat{\Psi}$. The vector $\vec{e}_k$ is the unit vector along the $k$-th direction. The HEOM consist of an infinite number of equations. These equations can be truncated at finite order when $\sum_{k} n_{k}$ first exceed a properly chosen large value $N$.\cite{IshizakiJPSJ05}

\subsection{Flux-flux correlation function: chemical reaction rate}
\label{sec:linearrate}

A chemical reaction process is typically characterized by a rate constant defined by the flux-flux correlation (FFC) function. \cite{YamotoJCP1960,TroMillFDCS1987,TopMak1994JCP,Miller1983}
In this paper, we consider the time dependent rate constant which is defined in terms of the canonical correlation function as \cite{YamotoJCP1960}
\begin{align}
k(t) &=\frac{1}{\beta}\int_0^\beta {d\lambda} \,
{\rm Tr}\left\{ \hat{\rho}_{\rm tot}^{\rm eq} e^{\lambda \hat{H}_{\rm tot}} \dot{{\hat \theta}}_R e^{-\lambda \hat{H}_{\rm tot}}  \dot{{\hat \theta}}_R(t)\right\} \nonumber \\
&\equiv \int_0^t dt  \left\langle \dot{{\hat \theta}}_R; \dot{{\hat \theta}}_R(t) \right\rangle ,
\label{eq:FFrateCoef}
\end{align}
where $\hat \theta_R$ is the projection operator introduced to evaluate the population of the reactant, and $ \dot{\hat{\theta}}_R=[ \hat{\theta}_R,\hat{H}_{\rm tot}]/i\hbar$. Typically, $\hat \theta_R$ is chosen to be the Heaviside step function of the proton coordinate $q$.\cite{YamotoJCP1960} By using Kubo's identity, Eq.\eqref{eq:FFrateCoef} can be recast as
\begin{align}
k(t)= \frac{i}{\hbar}\int_0^t dt  {\rm Tr} \left\{ \dot{\hat{\theta}}_R \hat{ \mathcal{G}}(t)
{{\hat{\theta}}_R}^{\times} \hat{\rho}_{\rm tot}^{\rm eq}\right\},
\label{eq:raterecast}
\end{align}
where $\hat{\mathcal{G}}(t)$ is the Liouville space time propagator, which is evaluated from Eq.\eqref{eq:HEOM}, and $\hat\rho_{\rm tot}^{\rm eq}(t)$ is the equilibrium density operator of total system, which can be prepared by integrating Eq.\eqref{eq:HEOM} over sufficiently long time from the factorized initial state, $\hat \rho_{\rm tot}(-\infty)=\hat\rho(0){\otimes} \hat\rho_{\rm B}^{\rm eq}$. In the HEOM formalism, the correlated (un-factorized) thermal equilibrium state can be set by using this steady state solution of the hierarchal elements.\cite{Tanimura2014,Tanimura2015, Shi2015CorretedInitial}

The right-hand side of Eq.\eqref{eq:raterecast} can be read as follows. 
At $t=0$, the system in the thermal equilibrium state is excited by the first interaction ${{\hat{\theta}}_R}^{\times}$, and propagated over a time $t$ by $\hat{\mathcal{G}}(t)$ .The first-order response function is then calculated from the expectation value of $\dot{\hat{\theta}}_R$. This is the numerical simulation for linear response measurement.\cite{TanimuraJCP92,KuboJPSJ1957} The rate constant is evaluated as $k_{\rm cnt}= k(+\infty)$. 

\section{Results and Discussion}
\label{sec:result}

\begin{figure}[b]
\includegraphics[width=0.8\textwidth]{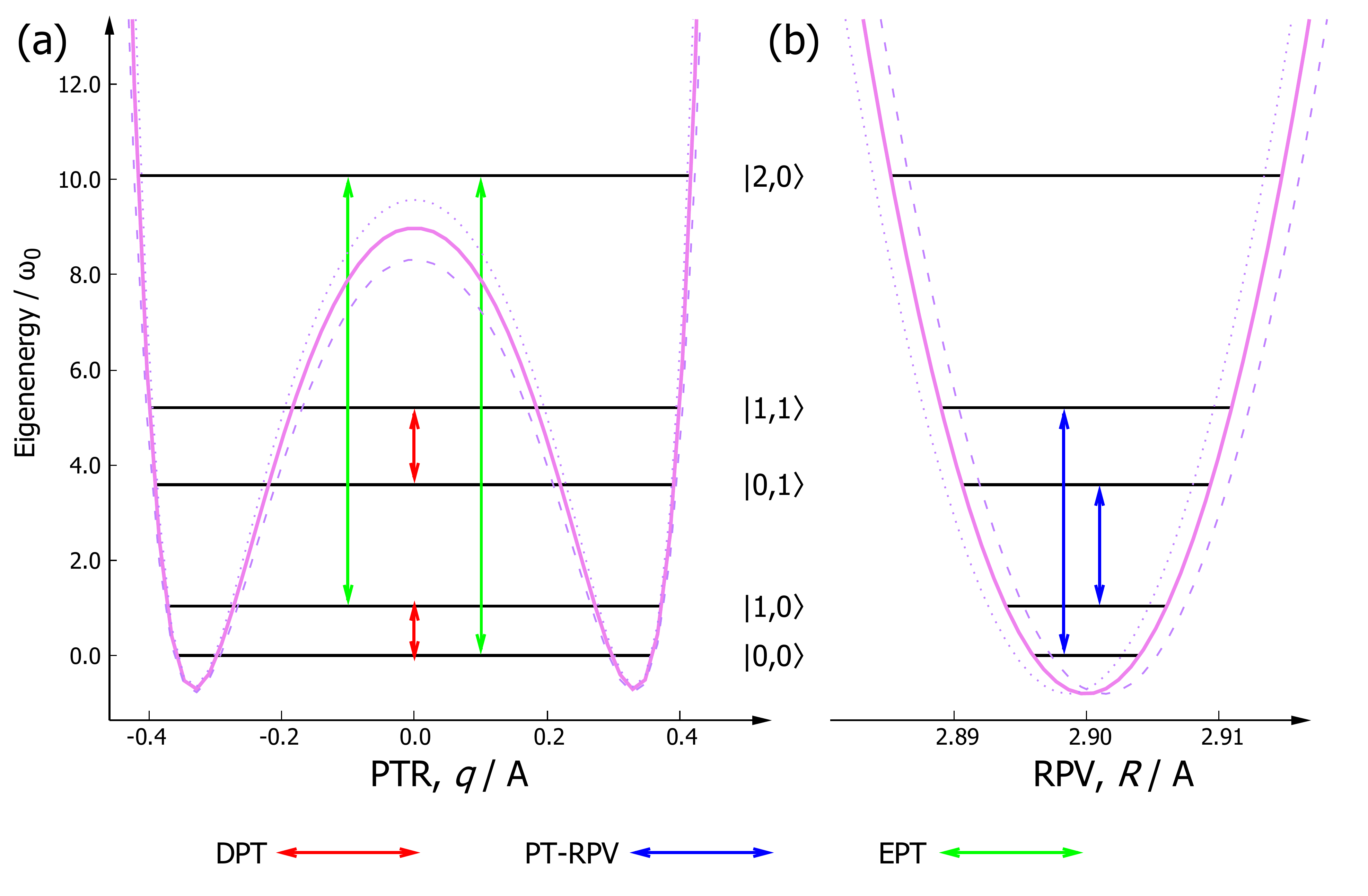}
\caption{Energy eigenstates $\lvert m, n \rangle$ and the PES are schematically depicted in (a) the $q$ direction for fixed $R=2.9$ and (b) the $R$ directions for the fixed $q=0.35$.  The solid violet curve represents the unperturbed PES described by $U_0 (\hat q, \hat R)$, while the dotted and dashed curves represent the PES under the positive and negative bath perturbations. The EL system-bath interaction causes frequency modulation in the $q$ direction and the shift of potential in the $R$ direction. Thus, we observe vibrational dephasing in the PTR mode, while we observe population relaxation in the RPV mode. The DPT, PT-RPV, and EPT transitions are illustrated by the red, blue and green arrows, respectively.} 
\label{fig.pjc}     
\end{figure}

The system parameter adopted in this work are summarized in Table. \ref{table.parameter} and are representative of typical hydrogen-bridged systems.
The system Hamiltonian is expressed in the energy eigenstate representation. Each eigenstate is labelled as $\lvert m, n \rangle$, where $m$ and $n$ represent an eigen-number of the PTR and RPV modes respectively. This labeling is certainly meaningful for low lying vibrational states while it can be problematic for higher energy excited states where the two modes can mix significantly. Furthermore, it enables us to clearly distinguish different paths of the proton-transfer process. The procedure for labeling is defined in the Appendix \ref{sec:Uqr}.
In order to numerically solve the HEOM, we employ the fourth-order Runge-Kutta method with a time step of $\delta t=0.01/\omega_0$, where $\omega_0 = $110 cm$^{-1}$ is the lowest excitation energy of system.  (See Appendix \ref{sec:Uqr}).
The depth and length of the hierarchy is chosen to be $N=20 \sim 50$ and $K=0 \sim 1$ with respect to different $\gamma$, and the lowest $12$ eigenstates are utilized to describe the system. The accuracy of the calculation is maintained by increasing the values of $N$ and $K$ till numerical convergece. 

\subsection{Effective coupling strength and flux-flux correlation function}

\begin{figure}[t]
\includegraphics[width=0.9\textwidth]{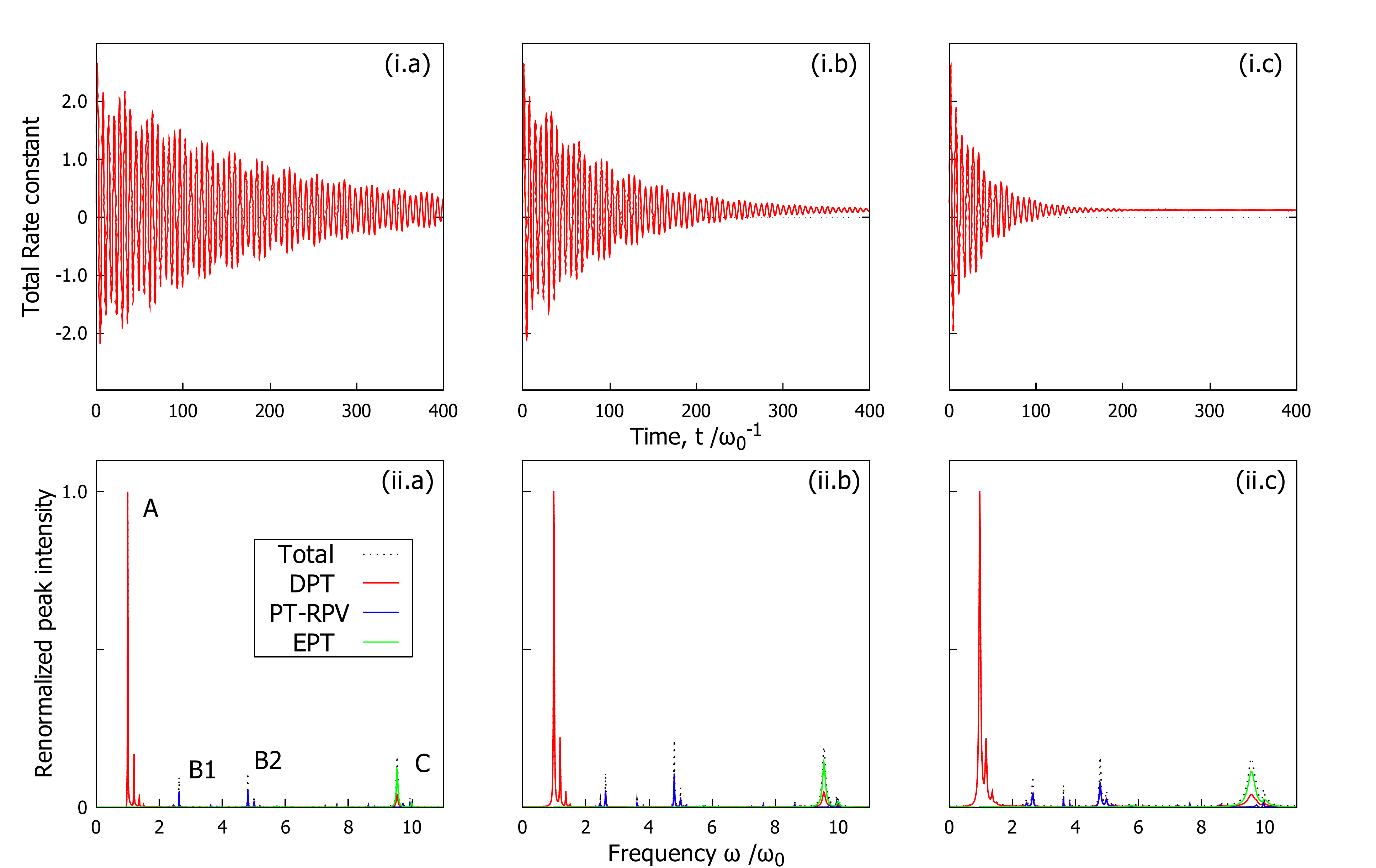}
\caption{The time evolutions of the flux-flux correlation (FFC) function, $k(t)$, are presented in the (i) real-time and (ii) Fourier space for the (a) weak ($\zeta '=0.1 \times 10^{-2} \omega_0$), (b) intermediate ($\zeta ' = 0.2 \times 10^{-2} \omega_0$), and (c) strong ($\zeta ' = 0.5 \times 10^{-2} \omega_0$) system-bath coupling cases, respectively. Here we set $T=$ 300K and $\gamma = \omega_0$.}
\label{fig.ffc}
\end{figure}

First we note that when the inverse noise correlation time, $\gamma$, becomes large, the effective coupling strength increases, even if $\zeta$ is fixed, because the bath can interact with the system multiple times when the correlation time is short.\cite{TanimuraJCP92} For the Drude SDF, Eq.\eqref{eq:JDrude}, the effective coupling strength is expressed as $\zeta'=\zeta\gamma^{2}/(\gamma^2+\omega_0^2)$. In the following, we employ $\zeta'$ to identify the pure non-Markovian effects and the pure non-perturbative effects separately. 

Fig. \ref{fig.ffc} illustrates the time evolution of the FFC function, $k(t)$, in (i) real time and (ii) Fourier space for the (a) weak, (b) intermediate, and (c) strong system-bath coupling cases for the fixed temperature ${T=300\rm{K}~(\beta \hbar \omega_{0} = 0.53)}$, and the inverse noise correlation time $\gamma=\omega_0$. 
In the present deep tunneling case, the time evolution of the FFC is highly oscillatory, as the system dynamics characterized by the various excitation energies, in comparison with the semi-classical case described by a Brownian based model with the LL interaction. In order to investigate the role of the  transitions among the system states, we show the Fourier transformation of the signal in Fig. \ref{fig.ffc}(ii). 
At $T=$ 300K ($\omega_{300{\rm K}}\approx 200$cm$^{-1}$) the system is initially in the low energy states, yet high frequency transition peaks in Fig. \ref{fig.ffc}(ii) are observed, due to the excitations induced by the operator $\theta_R$.

Using Table \ref{table.eigenstates} in Appendix \ref{sec:Uqr}, we can identify all  the transition peaks appearing in Figs. \ref{fig.ffc}(ii-a)- \ref{fig.ffc}(ii-c). The peak labeled as ``A''  corresponds to the transitions in the direct proton transfer (DPT) processes, ($\lvert 0,n \rangle\rightarrow\lvert 1,n \rangle$ for $n=1,2,...$)  and the peak labeled as ``C'' corresponds to the transitions in the excited proton transfer (EPT) processes ($\lvert 0,0\rangle \rightarrow\lvert 2,0\rangle$ and $\lvert 1,0\rangle \rightarrow\lvert 2,0\rangle$). 
Because the non-perturbative system-bath interaction can create a mixture of the PT and RPV vibrations, we observe the peaks that correspond to the PT-RPV transitions labeled as ``B1''($\lvert 1,n\rangle \rightarrow\lvert 0,n+1\rangle$) and ``B2''($\lvert 0,n\rangle \rightarrow\lvert 1,n+1\rangle $). A schematic view of these three transition processes is presented in Fig. \ref{fig.pjc}.  
It should be noted that these excitations decay toward the equilibrium state and the height and area of these transition peaks do not relate to the reaction rate that is determined near the equilibrium state. However, these profiles are helpful to investigate the dynamical aspect of the reaction processes. 

In comparison with the results obtained from a Brownian based model with the LL system-bath coupling,\cite{TanimuraPRA91,TanimuraJCP92,IshizakiJCP05,Shi2011PT,Shi2019ChargeCarrier} the present EL coupling case exhibits highly oscillatory FFC that are observed as the sharp peaks in Fig. \ref{fig.ffc}(ii-c), even in the strong system-bath coupling case. This distinguished feature is due to the lack of the LL component, $q  \sum_j g_j x_j$ for the PT mode in the EL system-bath coupling. Indeed, it has been shown that the SL interaction, $q^2  \sum_j g_j x_j$, contribute to the vibrational dephasing (or frequency fluctuation) rather than the population relaxation.\cite{TaniIshiACR09,SteffenTanimura00,TanimuraSteffen00,KatoTanimura02,KatoTanimura04,SakuraiJPCA11} Because the effect of the vibrational dephasing becomes larger for larger frequency modes (see Fig. \ref{fig.pjc}), the peak profiles of the high frequency EPT modes (``C'') are broadened, as illustrated in Fig. \ref{fig.ffc}(ii-c), while the profiles of the low frequency DPT modes (``A'') do not change even in the large system-bath coupling case. This indicates that the low frequency modes decay slower than the high frequency modes, as can be observed in Fig. \ref{fig.ffc}(i). 

\subsection{Temperature effects: role of fluctuations}

\begin{figure}[b]
\includegraphics[width=0.9\textwidth]{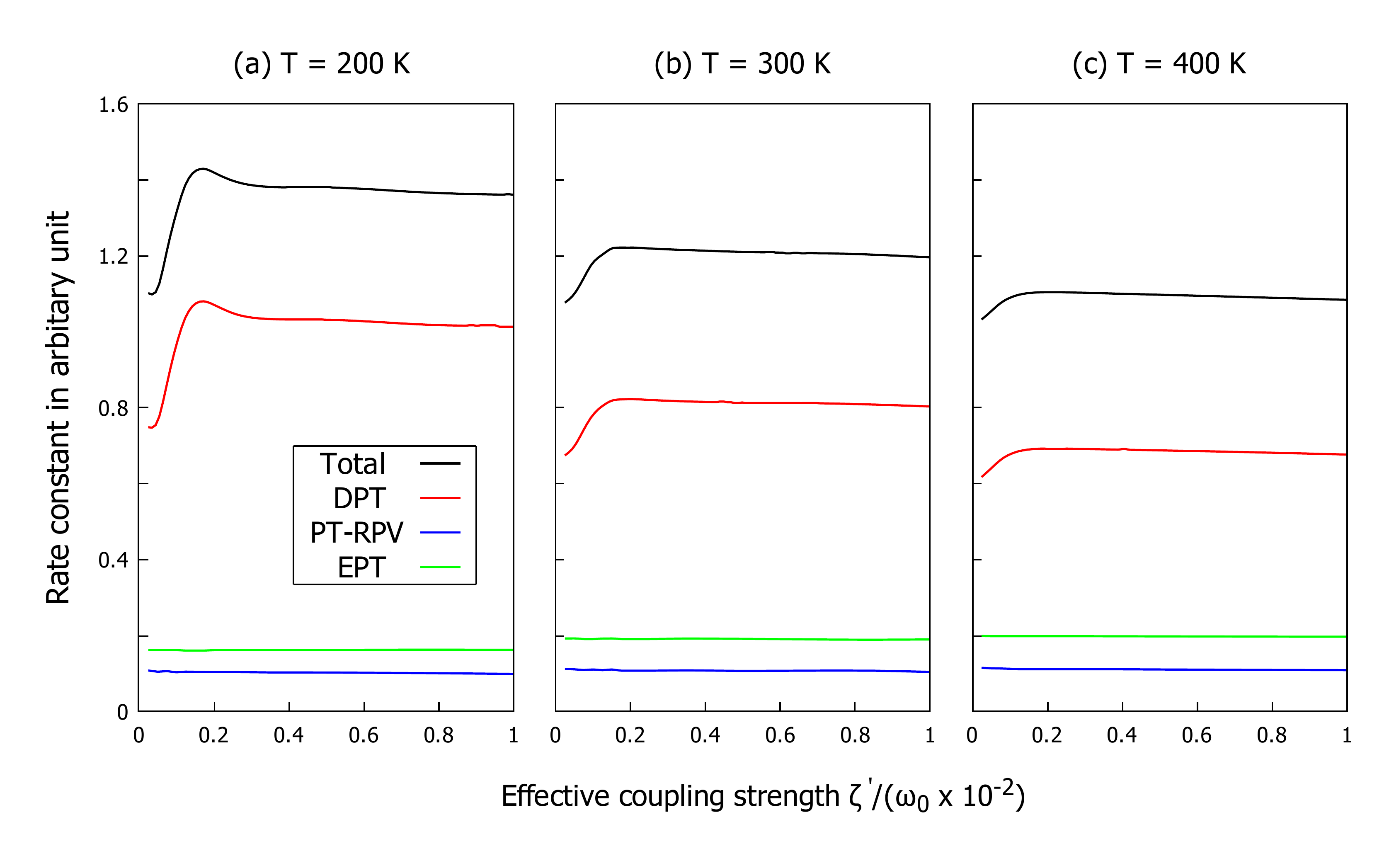}
\caption{The rate constants $k_{\rm{cnt}}$ are plotted as a function of the effective coupling strength $\zeta'$ at different temperature $T:$ (a)~$200\rm{K}$, (b)~$300\rm{K}$, and (c)~$400\rm{K}$. The contributions from the direct proton transfer (DPT), excited proton transfer (EPT), and the PT-RPV processes are also depicted in the same colored curves as Fig. \ref{fig.ffc}.(ii). The inverse correlation time is $\gamma = \omega_0$. The abscissa has been scaled by a factor $0.01\omega_0$ only to improve the readability of the figure.}
\label{fig.betahbar}
\end{figure}

Next we consider the temperature dependence of the rate constant $k_{\rm cnt}$ for a fixed inverse noise correlation time $\gamma=\omega_0$. In Fig. \ref{fig.betahbar}, we compare the rate constant as a function of coupling strength for different temperatures $T$: (a)~$200{\rm K}~{(\beta \hbar \omega_{0} = 0.78)}$, (b)~$300{\rm K}$, and (c)~$400\rm{K}~{(\beta \hbar \omega_{0} = 0.40)}$. The thermal excitation energy in this temperature regime is around $130 \sim 270 {\rm cm}^{-1}$. Thus the DPT modes whose excitation energy is approximately $\omega=110 \sim 150{\rm cm}^{-1}$, are thermally well activated and provide the main contribution to the reaction rate. In order to illustrate this point, we separately plot the contribution from the DPT, PT-RPV and EPT processes, which correspond to the peaks labeled by ``A",``B1'' and ``B2'', and ``C'' in Fig. \ref{fig.ffc}(ii), respectively. (see Appendix \ref{sec:pjcop}). Because the EL system-bath interaction creates transitions among the system states in a complex manner DPT, EPT and PT-RPV processes are highly entangled, as illustrated in the peaks in Fig. \ref{fig.ffc}(ii). The EPT and PT-RPV contributions are relatively small regardless of the coupling strength due to the large excitation energies ($\omega=500 \sim 1000{\rm cm}^{-1}$). 

In the classical and quantum rate theory (Kramers theory) developed on the basis of the LL Brownian based model, the turn-over feature from the energy controlled regime to the diffusion regime has been observed.\cite{Wolyness1981,Miller1983,Miller1989,HanggiRMP90,Shi2011PT,Shi2019ChargeCarrier,TanimuraPRA91,TanimuraJCP92} In the present case, while the reaction rate increases as $\zeta '$ increases in the small coupling strength region, we cannot observe the clear turnover feature for large $\zeta '$. This difference is due to the EL system-bath interaction. As discussed in Sec. \ref{sec:SB}, the major effects of the EL interaction with the PT mode is not population relaxation but vibrational dephasing. Thus, even for large $\zeta'$, the energy relaxation of the system states is slow, while non-dissipative vibrational modulation (or fluctuation) becomes larger. As a result, the reaction rate exhibits a plateau like behavior for large $\zeta'$, while the rate predicted from the LL Brownian model decreases, which has been explained by the quantum version of the Kramers turn-over theory.\cite{TanimuraPRA91,TanimuraJCP92}

The distinct difference between the LL Brownian case and the present case is observed in the high temperature case. In the framework of the LL Brownian theory, the reaction rate becomes larger for higher temperature, because the thermal activation process of the system enhances the reaction rate. In the present case, however, the reaction rate becomes smaller for higher temperature. This is because the DPT process, which plays a dominant role in the PT reaction process, is suppressed due to the large vibrational dephasing  effects arising from the thermal fluctuation, which depends on the temperature. For the EPT and PT-RPV processes, the excitation energies are much larger than the thermal excitation energy, and their contributions are small and do not change regardless of the temperature. 
Especially, when $T\to 0\rm{K}$, the excitation of RPV mode will be suppressed, and the 2D system will be similar to the traditional 1D case (coupled to a heat bath at 0 K).

\subsection{Non-Markovian and Markovian thermal effects}

\begin{figure}[t]
\includegraphics[width=0.9\textwidth]{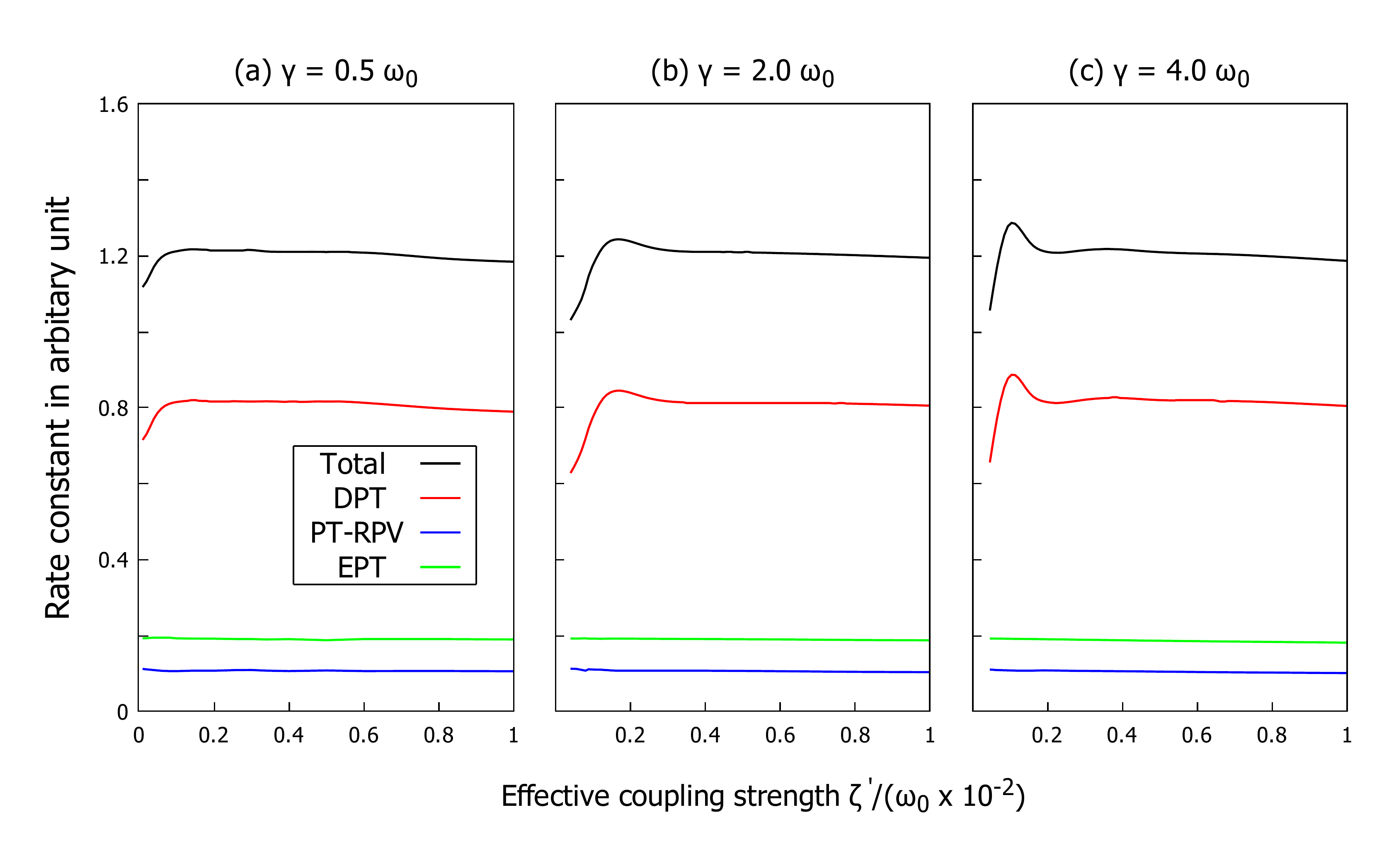}
\caption{The rate constant $k_{\rm{cnt}}$ as a function of $\zeta '$ for different inverse correlation time (a) $\gamma= 0.5 \omega_0$, (b) $2.0 \omega_0$, and $ 4.0 \omega_0$ at the fixed temperature $T=300\rm{K}$. The case for $\gamma= 1.0 \omega_0$ at $T=300\rm{K}$ is presented in Fig. \ref{fig.betahbar}(b). The contributions from the DPT, EPT, and PT-RPV processes are also depicted as the colored curves.}
\label{fig.gamma}
\end{figure} 

Finally, we consider the effect bath correlation time on the rate constant to investigate the role of non-Markovian and Markovian dynamics for the EL system-bath interaction. In Fig. \ref{fig.gamma}, we present the rate constant as a function of $\zeta'$ for different $\gamma$. The case $\gamma=\omega_0$ is presented in Fig. \ref{fig.betahbar} (b). For the slow modulation case in Fig. \ref{fig.gamma}(a), the reaction rate does not depend on the coupling strength except for the region in the very small $\zeta '$. This is due to the slow bath modulation in the EL coupling case, in which the system evolves with time under the perturbed static potential.

For the intermediate modulation cases in Fig. \ref{fig.betahbar}(b) and Fig. \ref{fig.gamma}(b), the rate increases with $\zeta'$ in the very weak coupling region. This is because the thermal activation process involved in the RPV-bath interaction, whose leading order agrees with the LL interaction (i.e. $R \sum_j g_j x_j$) takes place. This can be verified by changing the number of the RPV states, $n$, involved in the calculation. (see Appendix \ref{sec:RPV}). For the large $\zeta '$, vibrational dephasing, arising from the SL component of the EL interaction in the $q$ direction, plays a dominant role instead of the LL component of the RPV mode in the $R$ direction. Thus we observe the turn-over feature for small $\zeta '$ as the quantum Kramers theory predicted, while the plateau like feature of the reaction rate for large $\zeta'$ is observed due to the vibrational dephasing in the PTR mode. Regardless of $\gamma$, the contribution from the PT-RPV and EPT processes do not change, because the vibrational dephasing plays a minor role for the relaxation process. 

In the case of very fast modulation (the Markovian limit or the motional narrow limit) in Fig. \ref{fig.gamma}(c), the maximum peak of the turn-over feature becomes more prominent due to the contribution from the higher RPV states in the relatively small coupling region. Regardless of the values of $\gamma$, the final value of the plateau like feature is almost the same. This indicates that the vibrational dephasing effect arising from the SL component is determined by $\zeta '$, instead of $\zeta$ or $\gamma$. 

\section{Conclusion}
\label{sec:Conclusion}
In this paper, we introduced a model described by a two-dimesional potential surface with a realistic EL system-bath interaction for the investigation of the proton transfer reaction rate. The distinct feature of the present mode arises from the EL interaction that is characterized by the SL system-bath interaction for the PTR mode and the LL system-bath interaction for the RPV mode. The LL interaction contribute to the population relaxation, while the SL interaction contribute to the vibrational dephasing. The interplay between the PTR mode and RPV mode through the EL interaction exhibit distinctive features of the reaction process: in comparison with  previous studies based on the Brownian LL model, the turn-over feature of the reaction rate as a function of the system-bath coupling is suppressed, because the vibrational dephasing does not contribute the population relaxation so much. Most prominently, the reaction rate decreases with the temperature increases, which is the opposite to the prediction from the Kramers theory.
While the lack of a linear coupling between the proton coordinate and the bath degrees of freedom might seem a strong assumption we point out that in our two-dimensional model a significant the potential $U(q,R)$ explicitly includes  the dependence of the equilibrium position of the proton coordinate on the RPV coordinate which is certainly a fundamental aspect to take into account to describe any proton transfer process.

Although the present investigations are limited to a specific model, we believe that the applicability of our finding for the exponential-linear system-bath coupling is wider as it includes most of the fundamental types of system-bath interactions. Proton-Coupled Electron Transfer (PCET), Electron-Driven Proton Transfer (EDPT) mechanisms are also expected to show the type of system-bath interactions described here.
 Moreover, we can easily extend the present model to employ different system-bath coupling in the framework of the HEOM theory.

\begin{acknowledgments}
The financial support from The Kyoto University Foundation is acknowledged.
\end{acknowledgments}

\section*{data availability}
The data that support the findings of this study are available from the corresponding author upon reasonable request.

\appendix
\section{Energy eigenstate representation of the PT system}
\label{sec:Uqr}
In this paper, we describe the system Hamiltonian and interaction function in terms of system eigenstates, calculated from Schr\"{o}dinger equation,
\begin{align}
\left[ -\frac{1}{2} \left \{ \left ( \frac{1}{m}+\frac{1}{M}\right)  \frac{\partial^2}{\partial x^2} 
+ \frac{2}{M}  \frac{\partial^2}{\partial R^2}
+\frac{2}{M} \frac{\partial^2}{\partial x \partial R} \right \}  + U(\hat q, \hat R)  \right] \Phi \, = \, E \Phi \, .
\label{eq:schorodinger}
\end{align}
The parameters we used are listed in Table \ref{table.parameter}, based on the data of ref.  \onlinecite{SatoIwata}.

\begin{table}[b]
\caption{System parameters}
\begin{tabular}{c|c|c}
\hline
\hline
Mass & $M$ &100 a.m.u \\ 
	   & $m$ & 1 a.m.u \\
\hline
RPV  mode & $D_k$ & 303435 ${\rm cm}^{-1} {\rm \AA}^{-2}$ \\
 & $R_e$ & 3.0 ${\rm \AA}$ \\
\hline
Double well Morse & $D$ & 33715 ${\rm cm}^{-1}$ \\
  & $r_e$ & 1.0~${\rm \AA}$ \\
  & $\alpha_0$ & 2.0~${\rm \AA}$ \\
\hline
\hline
\end{tabular}
\label{table.parameter}
\end{table}

The quantum numbers $m$ and $n$ labeling the eigenstates of the system are determined from the number of nodes along the $q$ and $R$ directions respectively. 
The eigen-energies for the wave functions $\lvert m, n \rangle$  are presented in in Table \ref{table.eigenstates}.
The states $\lvert 0,n\rangle$ and $\lvert 1, n\rangle $ are the typical symmetric and asymmetric tunneling states, which is similar to the one-dimensional case. The states $\lvert 2,n\rangle$ and above are the delocalized proton excited states. The energy of first eigenstate $\lvert 0,0\rangle$ is set to be zero. The characteristic frequency $\omega_0$ is chosen to be the energy difference between state $\lvert 0,0 \rangle$ and $\lvert 1,0 \rangle$, which is near $110 {\rm cm}^{-1}$, and is used as the unit during the calculation. 

\begin{table}[h]
\caption{Eigen energy for the wave function $\lvert m,n \rangle$ for the PTR and RPV eigen states, $m$ and $n$.}
\begin{tabular}{c|c|c}
\hline
\hline
No. &  $\lvert m,n \rangle $ & Eigen energy/$\omega_0$ \\ 
\hline
1	& $\lvert 0,0 \rangle $	& 0.00 \\
\hline
2	& $\lvert 1,0 \rangle $	& 1.00 \\
\hline
3	& $\lvert 0,1 \rangle $	& 3.61 \\
\hline
4	& $\lvert 1,1 \rangle $	& 4.81 \\
\hline
5	& $\lvert 0,2 \rangle $	& 7.26 \\
\hline
6	& $\lvert 1,2 \rangle $	& 8.63 \\
\hline
7	& $\lvert 2,0 \rangle $	& 10.52 \\
\hline
8	& $\lvert 0,3 \rangle $	& 10.94 \\
\hline
9	& $\lvert 1,3 \rangle $	& 12.44 \\
\hline
10	& $\lvert 0,4 \rangle $	& 14.53 \\
\hline
11	& $\lvert 2,1 \rangle $	& 14.79 \\
\hline
12	& $\lvert 1,4 \rangle $	& 16.25 \\
\hline
\hline
\end{tabular}
\label{table.eigenstates}
\end{table}

\section{ Projection operator }
\label{sec:pjcop}

We consider the projection operator for the population of the acceptor state for the $n$th eigenstate of the RPV mode $\hat \theta_R^{(n)}$.  The other projection operator for the donor state is given by  $\hat \theta_L^{(n)} = 1 - \hat \theta_R^{(n)}$.
The operator $\hat \theta_R^{(n)}$ is defined by
\begin{align}
\hat \theta_R^{(n)} = {\hat h}(x)~( {\hat \vartheta}^{n} + {\hat \vartheta}^{E}),
\label{eq:projecop}
\end{align}  
where  $\hat h(x)$ is the step function and ${\hat \vartheta}_{n}$ and $ {\hat \vartheta}_{E}$ are the projection operators for $\lvert m,n \rangle$ defined by
\begin{align}
{\hat \vartheta}^{n^{,}} ~\lvert m,n \rangle =
\begin{cases}
~\lvert m,n \rangle & \mbox{if $m=0$ or 1, $n^{'}=n$} \\
~0 & \mbox{if $m=0$ or 1, $n^{'}\ne n$}
\end{cases}
\end{align}
and
\begin{align}
{\hat \vartheta}^{E} ~\lvert m,n \rangle = 
\begin{cases}
~0 & \mbox{if $m=0$ or 1} \\
~\lvert m,n \rangle & \mbox{if $m\geqq 2$}
\end{cases}
\end{align}
Using $\hat \theta_L^{(n)}$ and $\hat \theta_R^{(n)}$, we can separately evaluate the contribution form the donor and acceptor states for different by $n$. Then, we can identify the contribution from DPT processes, ($\lvert 0,n \rangle\rightarrow\lvert 1,n \rangle$ for $n=1,2,...$), PT-RPV processes ($\lvert 1,n \rangle \rightarrow\lvert 0,n+1\rangle$ and $\lvert 0,n \rangle \rightarrow\lvert 1,n+1\rangle$) and EPT processes ($\lvert 0,0\rangle \rightarrow\lvert 2,0\rangle$ and $\lvert 1,0\rangle \rightarrow\lvert 2,0\rangle$).

\section{Role of the RPV states}
\label{sec:RPV}
\begin{figure}[t]
\includegraphics[width=0.8\textwidth]{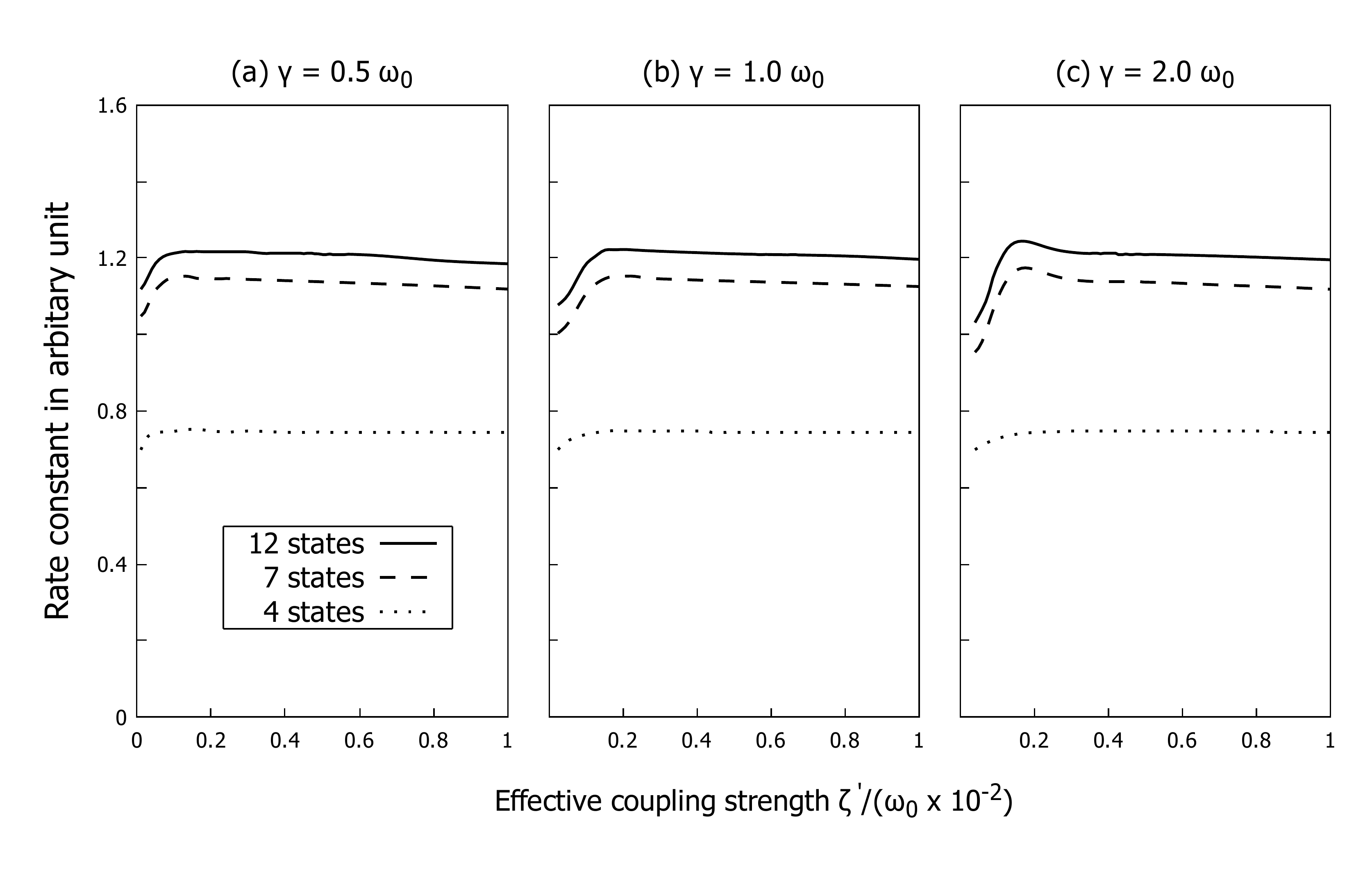}
\caption{The rate constant $k_{\rm{cnt}}$ for different RPV states described by $n$. The result for four, seven, and twelve eigenstates (the same condition in Fig.\ref{fig.gamma}) are depicted in dashed, dotted and solid curves. The temperature is fixed as $T=300\rm{K}$. }
\label{fig.nrpv}
\end{figure}

In order to investigate a role of the RPV states $n$, we calculate the reaction rate with changing the number of system eigenstates. Fig. \ref{fig.nrpv} shows the calculated results of the reaction rate for the cases of the four  ($n=1$), seven ($n=2$), and twelve ($n=4$) eigenstates. 
When $n$ is small, the result exhibit similar profile as in Fig. \ref{fig.gamma}(a). The rate calculated with 4 eigenstates is significantly smaller because the contribution from the EPT process is not involved.  This RPV-bath LL interaction is prominent in the small $\zeta'$ region that leads the turn-over feature, as quantum Kramers theory predicted. In the large $\zeta'$ region, the vibrational dephasing, which is described by the SL interaction in the PTR mode, becomes dominant and thus the reaction rate shows the plateau like feature.

\end{document}